\nonstopmode
\documentclass[aip,pop,reprint,superscriptaddress]{revtex4-2}

\usepackage{graphicx}
\usepackage{natbib}
\usepackage{amsmath,amssymb,amsfonts}
\usepackage[usenames,dvipsnames]{color}
\usepackage{subcaption}
\usepackage[normalem]{ulem}

\begin{document}

\title{NUMERICAL STUDY OF NON-GYROTROPIC ELECTRON PRESSURE EFFECTS IN COLLISIONLESS MAGNETIC RECONNECTION}

\author{A.~Sladkov}\affiliation{Institute of Applied Physics, 46 Ulyanov Street, 603950 Nizhny Novgorod, Russia}
\affiliation{Lab. Phys. Plasmas, Sorbonne Universit\'es, 4 place Jussieu, F-75252 Paris, France}
\author{R.~Smets} \affiliation{Lab. Phys. Plasmas, Sorbonne Universit\'es, 4 place Jussieu, F-75252 Paris, France}
\author{N.~Aunai} \affiliation{Lab. Phys. Plasmas, Sorbonne Universit\'es, 4 place Jussieu, F-75252 Paris, France}
\author{A.~Korzhimanov}\affiliation{Institute of Applied Physics, 46 Ulyanov Street, 603950 Nizhny Novgorod, Russia}

\begin{abstract}
We investigate the time evolution of the six-component electron pressure tensor in a hybrid code studying consequences for the two-dimensional reconnection process in an initially perturbed Harris sheet. We put forward that two tensor components (a diagonal and a non-diagonal one) grow in an unstable way unless an isotropization operator is considered. This isotropization term is physically associated with an electron heat flux. As a consequence, we put forward that an enhanced value of a diagonal component is observed in the very middle of field reversal at sub-ion scale. Because of the increase of the kinetic pressure, the magnetic field is decreased in this electron layer, hence increasing the associated out-of-plane current at its edges and leading to its bifurcation. The bifurcation mechanism is based on the presence of electron pressure anisotropy, related to the gradient of inflow electron bulk velocity. The gradient in the inflow direction of the enhanced diagonal electron pressure tensor component results in the deceleration of the ions entering the X-point region. We suggest that bifurcated current sheets resulting from the anisotropies/agyrotropies of the six-component electron pressure tensor correspond to smaller reconnection rates comparing to non-bifurcated ones.

\end{abstract}

\maketitle

\section{Introduction} \label{sec:intro}

Magnetic reconnection is well known to be an ubiquitous plasma process, occurring in laboratory\cite{taylor1986}, space\cite{dungey1961} and astrophysical\cite{degouveia2010} plasmas. It has been widely studied as it is an important transport and energy conversion process. While in nature reconnection could be very unsteady, many numerical simulations of a single X-point developing in a two-dimensional Harris sheet suggested a nearly stationary process~\cite{wan2008} after the onset. It essentially results from the fixed-size computational domain, the limited number of particles, and the uniform upstream plasma. While artificial, such setups then allow to quantitatively investigate the reconnection process during this stationary phase.

The out-of-plane component of the electric field is crucial in studying two-dimensional magnetic reconnection as this term quantifies the rate at which magnetic flux is transported in and out from the reconnection region. Upstream of the reconnection region, the out-of-plane component of the electric field is essentially due to the $\mathbf V_i \times \mathbf B$ term, which is associated with the ions, drifting at velocity $\mathbf V_i$ in the magnetic field $\mathbf B$. Approaching the mid-plane of the current sheet, this term vanishes because both the magnetic field and the ion velocity vanish (the ions being demagnetized). It is known~\cite{birn2001geospace} that the out-of-plane electric field is sustained by the Hall term in the so-called ion diffusion region, while in the electron diffusion region, the electron pressure term\cite{hesse2011} seems to be dominant.

In his pioneering work, Vasyliunas\cite{vasyliunas1975} pointed out the importance of non-gyrotropic electron pressure tensor, as being able to account for the out-of-plane component of the electric field at the stagnation point. Different numerical works on symmetric reconnection have already shown that non-gyrotropic contribution to the reconnection electric field exceeds the corresponding bulk flow inertial contribution in the case without guide-field~\cite{kuznetsova1998, shay1998, cai2009}. The divergence of the electron pressure tensor is hence the main term for the out-of-plane electric field. Non-diagonal terms of the pressure tensor naturally result from the non-gyrotropy of the associated distribution function~\cite{hesse2011}. In full-particle-in-cell (PIC) codes, such effects are well-captured, but often at the cost of huge CPU time\cite{daughton2007} and/or unrealistic mass ratio and/or small physical domains. An alternative is to treat electrons as a fluid, like in hybrid-PIC or two-fluid codes.

In collisionless plasmas, the time evolution of the six-component electron pressure tensor contains three contributions. We adopt here the form introduced by Winske~\cite{winske1994} using index notation:

\begin{eqnarray}
\partial_t P_{ij} & = & - v_k\partial_k P_{ij} -P_{ij}\partial_k v_k  - P_{ik}\partial_k v_j - P_{jk}\partial_k v_i \nonumber\\
 & & -\frac{e}{m_e} [P_{ik}B_n\varepsilon_{knj} + P_{jk}B_n\varepsilon_{kni}] \nonumber\\
 & & - \partial_k Q_{ijk} \label{eqa}
\end{eqnarray}

where $\varepsilon_{knj}$ is the Levi-Civita symbol (We adopt the MKS unit system). Following the terminology of Hesse~\cite{hesse1995}, the driver term $[D]$ (the first line of right-hand side of Eq.~(\ref{eqa})) is associated with the transport of pressure at the electron fluid velocity $v$, the cyclotron term $[C]$ (second line) is associated with the rotation of pressure at the electron gyrofrequency, and the divergence of the electron heat flux (third line), is usually replaced by an isotropization term $[I]$. As a consequence, resolving Eq.~(\ref{eqa}) requires to correctly treat the time integration of the term $[C]$ at electron scale, but also to find an appropriate form for the term $[I]$. In this work, we call $X$ the field reversal direction, $Y$ the gradient direction for both magnetic field magnitude and density, and $Z$ the direction of the current associated with the field reversal.

The first numerical study~\cite{winske1994} investigating the time-evolution of the electron pressure tensor only considered the terms $[D]$ and $[I]$, assuming that the term $[C]$ also results in an isotropization process. This numerical study\cite{winske1994} already outlined the importance of the term $[I]$ in order to limit the size of the off-diagonal components of the electron pressure tensor. This study-case has been revisited~\cite{hesse1995} including the term $[C]$, numerically treated in an implicit way to remove the time step constraint resulting from the electron mass (the cyclotron term linearly depends on the electron gyrofrequency).

Previous works~\cite{le2010, ohia2012, egedal2013} used an anisotropic but gyrotropic electron pressure with a dedicated equation of state. This closure has been used in fluid and hybrid models and has been compared to full-PIC simulations in the strong guide-field case. As already pointed out~\cite{wang2015} while such a model successfully recovers structural features that are typically observed in full-PIC simulations, this closure is not applicable for anti-parallel reconnection.

In this study, we present an explicit method for the time integration of the evolution equation of the six-component electron pressure tensor in a hybrid code and discuss the importance of the electron heat flux acting as a regularizing term on each of the components of this tensor. In section \ref{sec:model}, we introduce the hybrid numerical model. In section \ref{sec:full}, we give details for the numerical integration of the electron pressure tensor. In section \ref{sec:runB}, we display the structures of the components of the pressure tensor and discuss their origin. In section \ref{sec:runCD}, we discuss the macroscopic consequences of regularizing the $P_{xx}$ and $P_{xy}$ components for the current sheet structure at electron scale. In section \ref{sec:kinetic}, we present a full-PIC simulation of the same initial value problem to compare the structure of the out-of-plane current density. Section \ref{sec:layer} is dedicated to discuss the consequences for the reconnection process.

\section{Numerical model} \label{sec:model}

In this work, electrons are described in a fluid way, by their three first moments: density, bulk velocity, and the six-component pressure tensor (pressure tensor being symmetric, by definition). Such an approximation is generally called a ten-moment description. To properly describe the decoupling of ions from the magnetic field in the ion diffusion region, we keep the ion description at the particle level. Hence, we use a hybrid code\cite{smets2011}.

The displacement current is neglected as the phase velocity of electromagnetic fluctuations is small compared to the speed of light. We then need an electron Ohm's law:

\begin{equation}
\mathbf E = -\mathbf V_i \times \mathbf B + \frac{1}{e n}(\mathbf J \times \mathbf B - \boldsymbol{\nabla} . \, \mathbf P_e) - \nu \Delta \mathbf J \label{ohm}
\end{equation}

In Eq.~(\ref{ohm}), $\mathbf V_i$ is the ion bulk velocity, $n$ is the electron density (equal to the total ion density by quasi-neutrality), $\mathbf J$ is the total current density equal to the curl of $\mathbf B$, $\mathbf P_e$ is the electron pressure tensor and $\nu$ is the hyper-viscosity. 

In our numerical model, the magnetic field and the density are normalized to their asymptotic values $B_0$ and $n_0$ respectively, lengths are normalized to the ion inertial length $d_0$ (calculated using the density $n_0$), times are normalized to the inverse of ion gyrofrequency $\Omega_0^{-1}$ (calculated using the magnetic field $B_0$) and velocities are normalized to the Alfv\'en velocity $V_0$ (calculated using $B_0$ and $n_0$). Mass and charge are normalized to the ion ones. The normalization of the other quantities follows from these ones. For the hyper-viscous term, we set $\nu = 10^{-3}$: while it provides a dissipation process at sub-ion scales, it is still smaller than the contribution of the electron pressure tensor, whatever the scale, for the reconnection process.

Electromagnetic fields are calculated on two staggered grids using a predictor-corrector scheme \cite{winske1986electromagnetic} in order to ensure a second-order scheme. The dynamics of the ions is solved using a first-order interpolation of the electromagnetic field\cite{boris1972proceedings}. At each time step, the particle moments, namely the density and the velocity, are computed in each cell using a first-order assignment function for each macro-particle. These ion moments are smoothed using a three-points stencil. Such smoothing helps to prevent the growth of small-scale electric fields in low-density areas and has limited consequences for the numerical diffusion processes.

We use a Harris sheet~\cite{harris1962harris} as an initial condition without guide-field for which $\mathbf B(y)=B_0\tanh(y/\lambda) \hat{\mathbf x}$ and $n(y)=n_0\cosh^{-2}(y/\lambda)$, and a half width $\lambda = 1$. This value is twice as large as the value used in previous works\cite{winske1994, hesse1994,  hesse1995, yin2001, kuznetsova2001}, such an increase helps to quench the tearing mode growth rate everywhere except at the initial X-point, then allowing to focus on the single X-point case study. A small initial magnetic perturbation\cite{zenitani2011} is superposed to the Harris equilibrium in the very middle of the box. Such a kinetic equilibrium is entirely defined by the temperature ratio $T_i/T_e$, equal to 1 in this study. Background ions have a uniform density $n_b$ (equal to 0.5 in this study) and the same temperature as foreground ions forming the current sheet.

The simulation domain is rectangular with a length $L_X = 102.4$ and a width $L_Y = 20.8$. We use a 1024$\times$208 grid corresponding to a mesh size equal to 0.1 in both directions. We use a single particle population with initially 40 million particles in the domain, meaning the number of particles per cell is initially greater than 160 in the lowest-density regions (namely the lobes). The time-step is $2\times10^{-4}$ in order to satisfy the CFL conditions for the fastest whistler modes. Each run lasts $t_{\max}$ = 100 $\Omega_0^{-1}$. We use periodic boundary conditions in  $X$ direction and perfect conducting boundaries in $Y$ direction as the magnetic field and density profiles are not periodic functions.

In the next section, we discuss the equation governing the time evolution of the electron pressure tensor, the importance of each of its terms, and the way these terms can be numerically computed.

\section{Six-component electron pressure tensor equation} \label{sec:full}

In most hybrid codes, the pressure is scalar (with a closure condition being either isothermal or adiabatic) which means any anisotropies and agyrotropies are neglected. Although simple, this hypothesis is generally hard to justify, and the consequences for the reconnection process are yet unclear. Especially in 2D current sheet, as it does not support the reconnection electric field. Furthermore, in the isothermal case, the electron temperature has to be uniform, insofar as any gradient would have no way to evolve through time under this hypothesis. As we use the six-component pressure tensor equation, further we discuss each term on the right-hand side of Eq.~(\ref{eqa}) to understand the underlying physical effects, as well as the associated numerical constraints.

The driver term is written

\begin{equation}
D(\mathbf P) = -\mathbf V_e . \boldsymbol{\nabla} \mathbf P - \mathbf P \boldsymbol{\nabla} . \mathbf V_e - \mathbf P . \boldsymbol{\nabla} \mathbf V_e - (\mathbf P . \boldsymbol{\nabla} \mathbf V_e)^T
\end{equation}

\noindent
where the electron bulk velocity $\mathbf V_e$ is expressed through the ion bulk velocity and the current density: $\mathbf V_e$ = $\mathbf V_i - \mathbf J/en$.

The four terms involved in $D$ can be split in three parts:
\begin{itemize}
\item
$D_{\mathrm{A}} = -\mathbf V_e . \boldsymbol{\nabla} \mathbf P$ is the advection of the electron pressure at the electron velocity. For such advection equation, we use the first-order upwind scheme which is well designed
\item
$D_{\mathrm{C}} = -\mathbf P \boldsymbol{\nabla} . \mathbf V_e$ is associated with the plasma compressibility. It is numerically integrated with an explicit first-order space-centered scheme
\item 
$D_{\mathrm{S}} = -\mathbf P . \boldsymbol{\nabla} \mathbf V_e -(\mathbf P . \boldsymbol{\nabla} \mathbf V_e)^T$ is the symmetric part of the driver term and is also integrated with a first-order space-centered scheme
\end{itemize}

It is worth noticing that no electron time scales are present in these terms as the electron mass is not involved. Hence, no particular numerical constraints arise because of the time integration of these terms. A second important remark is that both $D_{\mathrm{A}}$ and $D_{\mathrm{C}}$ act on a $P_{ij}$ component by only involving this same $P_{ij}$ component. This is not the case for the $D_{\mathrm{S}}$ term, for which the time evolution of diagonal terms will depend on the off-diagonal ones, and vice-versa. We will hence see that $D_{\mathrm{S}}$ is generally a leading term compared to $D_{\mathrm{A}}$ and $D_{\mathrm{C}}$.

The cyclotron term is written

\begin{equation}
C(\mathbf P) = -\frac{e}{m} [\mathbf P \times \mathbf B + (\mathbf P \times \mathbf B)^T]
\end{equation}

\noindent
This term depends on the electron mass. Consequently, a constraint arises on the time step needed to integrate it. In a previous study~\cite{hesse1995}, this term was integrated in an implicit way resulting in the unconditionally stable numerical scheme. In our study, we develop a different numerical scheme using a sub-cycling technique. We also have an implementation of the implicit one for the purpose of comparison. These two methods converge to the same results, with about 30\% of the total computational time saved using subcycling.

The last term of the right-hand side of Eq.~(\ref{eqa}) is the divergence of the electron heat flux. In the collisionless case, the heat flux expression is a complex problem, as the equation for its time evolution is a heavy third-order tensor equation involving the divergence of fourth-order moment of the electron fluid. Hammett and Perkins\cite{hammett1990} proposed an Ansatz form for the reduced heat flux which includes a set of free parameters. The parameters are defined such that the closure has a linear-response function very close to that of a collisionless, Maxwellian plasma. This so-called Landau-fluid closure is well designed in a spectral representation as it is local in wave-number space, however, it is non-local in real space. 

We then need to choose a model for the divergence of the heat flux and chose to write this isotropization term as

\begin{equation}
I(\mathbf P) = -\frac{1}{\tau} [\mathbf P - \frac{1}{3}\mathrm{Tr} (\mathbf P)\mathbf 1]
\end{equation}

where $\mathbf 1$ is the unit tensor, $\tau$ is the characteristic relaxation time scale. This isotropization term operates with the same efficiency for both diagonal and off-diagonal terms. We also emphasize that $I_{ij}$ only depends on this $ij$ component (no cross-terms) except through the implicit relation in trace operation for the diagonal components. This form of the isotropization term depends on the difference between the six-component pressure tensor and the third of its trace multiplied by the unit tensor, as in previous studies\cite{winske1994, hesse1994, wang2015}. This term hence only depends on a characteristic time $\tau$, while there are no clear arguments to justify how it should depend on a local value of the magnetic field or the thermal velocity. Eq.~(\ref{eqa}) then writes

\begin{eqnarray}
\partial_t \mathbf P & = & -\mathbf V_e . \boldsymbol{\nabla} \mathbf P - \mathbf P \boldsymbol{\nabla} . \mathbf V_e - \mathbf P . \boldsymbol{\nabla} \mathbf V_e - (\mathbf P . \boldsymbol{\nabla} \mathbf V_e)^T \nonumber \\
 & & -\frac{e}{m} [\mathbf P \times \mathbf B + (\mathbf P \times \mathbf B)^T] \nonumber \\
 & & -\frac{1}{\tau} [\mathbf P - \frac{1}{3}\mathrm{Tr} (\mathbf P)\mathbf 1] \label{eq_final}
\end{eqnarray}

The time step $\Delta t$ we use for the equations is small enough to resolve the gyro-motion of ions. But the implemented sub-cycling method for the explicit integration of the cyclotron term requires a smaller time step. Hence, defining $\mu = m_i/m_e$ as the ion to electron mass ratio, a time step $\Delta t/ \mu$ is small enough to properly treat the electron magnetization. In this study, we take $\mu = 100$, so that the electron inertial length is $d_e$ = 0.1 $d_i$. Because of time centering constraint, the electron pressure tensor is defined at half time step, advancing from $P_{n-1/2}$ to $P_{n+1/2}$ is done using $\mu$ sub-cycles with the algorithm (for $l \in [0, \mu-1]$)

\begin{equation}
\mathbf P_{n-\frac{1}{2}+\frac{l+1}{\mu}} = \mathbf P_{n-\frac{1}{2}+\frac{l}{\mu}} + \frac{\Delta t}{\mu} [C(\mathbf P_{n-\frac{1}{2}+\frac{l}{\mu}}) + D (\mathbf P_n) + I(\mathbf P_{n-\frac{1}{2}}) ]\label{integral}
\end{equation}

The driver term is hence extrapolated at the predictor phase and interpolated at the corrector one.

The plasma moments in such a modified Harris sheet are known to be subject to the growth of small-scale structures, being at a sub-ion scale. In order to make them clear, we tried to avoid, as much as possible, any kind of smoothing, generally needed in hybrid codes to lower the noise level on the ion moments. We then kept a modest smoothing on ion density and velocity using a three-points stencil, while the six-component electron pressure tensor was not smoothed. As a cost, to keep also smooth electron velocity derivatives in the driver part, we additionally smooth the current in the electron velocity expression. This makes an important difference with previous comparable hybrid simulations~\cite{winske1994, hesse1995, yin2001, kuznetsova2001}. As a consequence, we had to keep a small enough time step with $\Delta t = 2 \times 10^{-4}$. Furthermore, we were unable to reach a grid size smaller than $\Delta x = 0.1$ under these constraints. Although being restrictive, this framework then allowed to observe sub-ion scales structures resulting from the integration of Eq.~(\ref{eq_final}), and not observed before. The very low level of smoothing also forced us to preserve a large enough value of the density in the lobe, which is 0.5 in this study. Nevertheless, we verified in the isothermal case that varying the $n_b$ values between 0.2 and 0.5 had no consequences neither on the reconnection process nor on its rate.

In this study, we discuss four types of runs, depending on the specific closure equation which is used. These types are labeled with tags (ranging from $A$ to $D$), which meanings are given in Table \ref{tab}. By default, we use $\tau = 1$ for each case, unless the $\tau$ value is explicitly given. In case $D$, $P_{xx}$ and $P_{yy}$ are calculated using Eq.~(\ref{eq_final}) but then are set to half of their sum to be used for the next time step. The anisotropy between the $P_{xx}$ and $P_{yy}$ terms is hence artificially quenched, while these terms are calculated with their appropriate time evolution equation.

\begin{table}
\begin{center}
\begin{tabular}{cl}
\hline
\hline
Run tag ~~ & Closure equation\\
\hline
A & isothermal closure, scalar $P = n k_B T_0$\\
B & tensor $\mathbf P$ (with subcycling for $[C]$ term)\\
C & tensor $\mathbf P$ with $[I]$ only on $P_{xx}$ and $P_{xy}$\\
D & tensor $\mathbf P$ with $P_{xx} = P_{yy}$\\
\hline
\hline
\end{tabular}
\caption{Meaning of the four tags used for runs A, B, C and D.\label{tab}}
\end{center}
\end{table}

In the next section, we discuss results for run B, while runs C and D will be discussed and compared to run B in section \ref{sec:runCD}. Run A will be used for comparisons in both of sections \ref{sec:runB} and \ref{sec:runCD}.

\section{Phenomenology of the six-component electron pressure tensor}\label{sec:runB}

The structure of the diagonal components of the pressure tensor is discussed in the first part of this section, while the off-diagonal components are discussed in the second one.

\begin{figure}
\includegraphics{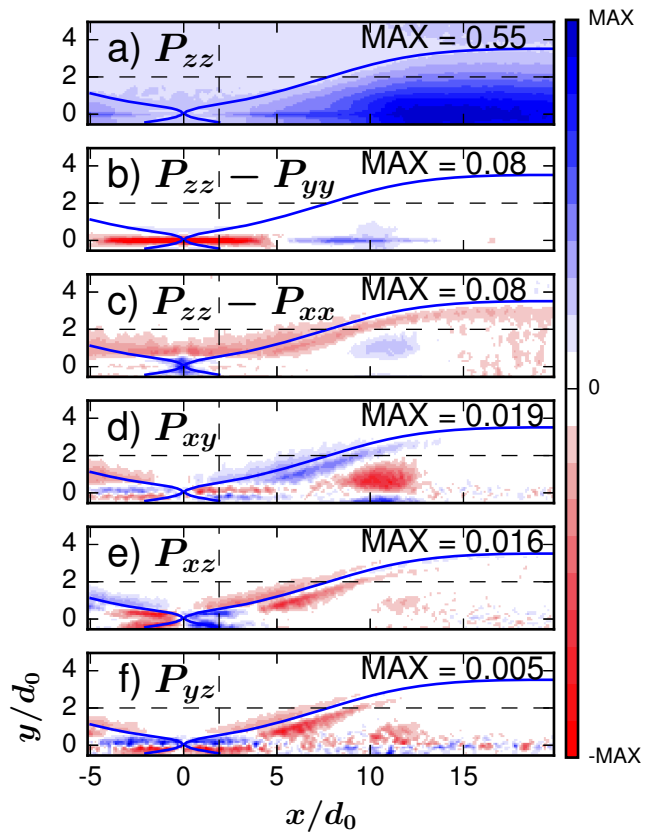}
\caption{ Run B. (a) Diagonal component  of the electron pressure tensor $P_{zz}$, (b) difference $P_{zz} - P_{yy}$, (c) difference $P_{zz} - P_{xx}$, (d)-(f) off-diagonal components at $t$ = 80 $\Omega_0^{-1}$. For each panel the respective maximum values are indicated in frame. The pressure is normalized to $P_0 = m_p d_0^{-1}\Omega_0^2$. The separatrices are indicated in blue lines. Dashed lines show cut locations used in analysis.} \label{fig:diag_offd_runB}
\end{figure}

Fig.~\ref{fig:diag_offd_runB} displays the electron pressure tensor components where we distinguish two regions: the separatrices and the electron layer, a region close to the mid-plane $Y$ = 0, at electron scale. In the very middle of the electron layer, the larger value of $P_{yy}$ compared to the two other diagonal components turns to be the parallel component of the pressure tensor as the only non-vanishing component of the magnetic field is in the $Y$ direction. In the anti-parallel reconnection case, such an increase of the parallel pressure at the sub-ion scale has already been reported both in fluid and full-PIC simulations\cite{wang2015} at very similar scales.

Panel (a) of Fig.~\ref{fig:diag_offd_runB} displays the $P_{zz}$ component of the pressure tensor and shows a very clear structure of the enhanced pressure in a thin and elongated layer close to the mid-plane $Y$ = 0. This electron layer has a length between 10 $d_0$ and 15 $d_0$ and has a thickness smaller than the ion inertial length (that is at electron scale). While this thickness also depends on the location and on the time, we call it electron layer as this thickness is always at electron scale. Such electron layer has already been observed in both full-PIC and two-fluid simulations\cite{wang2015}. Full-PIC simulations of the Harris-type magnetotail equilibrium showed that decreasing the electron mass down to realistic values resulted in decreasing the width of the non-gyrotropic region~\cite{liu2014}.

We will show later that the increased $P_{yy}$ value in the mid-plane is accompanied by a decreased density, therefore consistent with the heating of the electrons in full-PIC simulations~\cite{hesse2011}. Interestingly, the larger the isotropization term (replacing the divergence of the electron heat flux), the thinner the electron layer. It emphasizes that the electron layer results from the anisotropy and agyrotropy of the electron pressure. In our case, this non-gyrotropy arises from the interplay between all the terms in Eq.~(\ref{eq_final}). In a model accounting for the kinetic nature of the electrons, such non-gyrotropy in the pressure tensor would arise from the non-gyrotropies of the underlying electron distribution functions, which themselves result from the non-adiabatic electron motion in electron scale regions.

The $\tau$ value is the only external parameter controlling the efficiency of the isotropization process. We performed several simulations with $\tau$ ranging from an infinite value (that is without including the isotropization term) to $10^{-2}$ (that is strong isotropization on the electron scale). The former case is numerically unstable as both $P_{xx}$ and $P_{xy}$ grow in an unbounded way to unrealistic values. But this case is also unrealistic from a physical point of view, as there is no reason why the electron heat flux acting as a regularizing term should be canceled out. In the latter case the electron layer disappears, which hence put forward that the smaller $\tau$ (the larger isotropization), the stronger the dissipation of the electron layer (that is the thinner electron layer with a smaller magnitude of the electron current density). We kept $\tau=1$ to preserve the electron layer at the sub-ion scale, even if this value can be challenged. We use this value to put forward how the components of the pressure tensor evolve with time, and what are the underlying mechanisms.

Panels (b) and (c) of Fig.~\ref{fig:diag_offd_runB} display the difference of the $P_{zz}$ component with the two other diagonal components of this tensor. Two very clear structures appear that outlines the anisotropy of the electrons :
\begin{itemize}
\item in the electron layer, the $P_{yy}$ component is larger close to the $X$ point (starting at $X$ point until 5 $d_0$) and then gets smaller at the end of this electron layer
\item at the separatrices, the $P_{xx}$ component gets larger than $P_{zz}$ (and $P_{yy}$).
\end{itemize}
To clarify these features, we focus on the relative importance of the terms governing the time evolution of the six-component electron pressure tensor.

\begin{figure}
\includegraphics{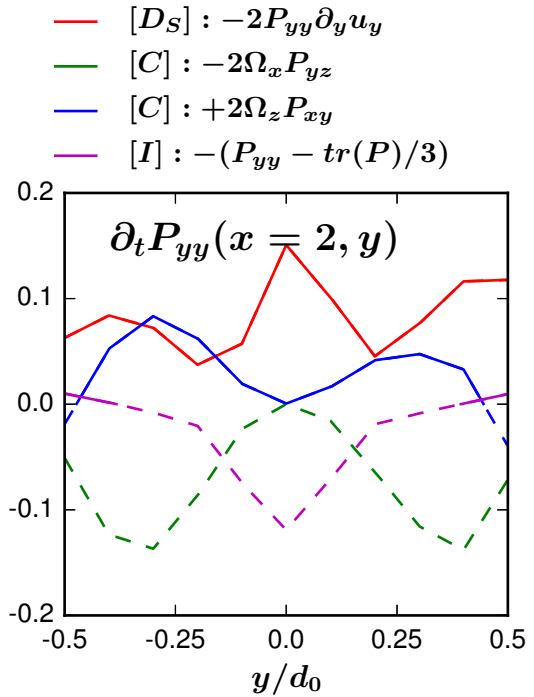}
\caption{Run B. Cut across the electron layer in $Y$ direction (at $X$ = 2 $d_0$, $t$ = 80 $\Omega_0^{-1}$) for the terms defined in Eq.~(\ref{eq_final}) for $P_{yy}$. For coloured lines, solid lines are used when the associated term increases the absolute value of pressure component, while dashed lines are used when this term decreases its absolute value.}\label{fig:pyy}
\end{figure}

Fig.~\ref{fig:pyy} displays the terms controlling the time evolution of the $P_{yy}$ component: the only non-negligible term in $[D_{\mathrm{S}}]$ (red), two terms in $[C]$ (green and blue) and $[I]$ (magenta). These values are averaged over 3 consecutive output steps (namely 79.9, 80.0, and 80.1) in order to mitigate their fluctuating nature. Solid lines are used when the associated term increases the absolute value of the pressure component while dashed lines are used when this term decreases its absolute value. In the electron layer, the picture is pretty clear, and exhibits a strong enhancement of the $P_{yy}$ component. As a consequence, an anisotropy is arising with $P_{yy}$ being larger than $P_{xx}$ and $P_{zz}$.

One understands that the origin of the $P_{yy}$ structure in the electron layer results from the component of $[D_{\mathrm{S}}]$ involving $P_{yy}$ itself and the gradient of the electron inflow bulk velocity $\partial_y V_y$. The growth of $P_{yy}$ by the $[D_{\mathrm{S}}]$ term is then self-fueled. Nonetheless, when we remove the isotropization term acting on $P_{yy}$ in run C, it is still limited by the term $[D_{\mathrm{A}}]$ (not shown here). It means that the growth of $P_{yy}$ in the electron layer is intrinsically finite. The term $[I]$ clearly acts in order to reduce the anisotropy.

\begin{figure}
\includegraphics{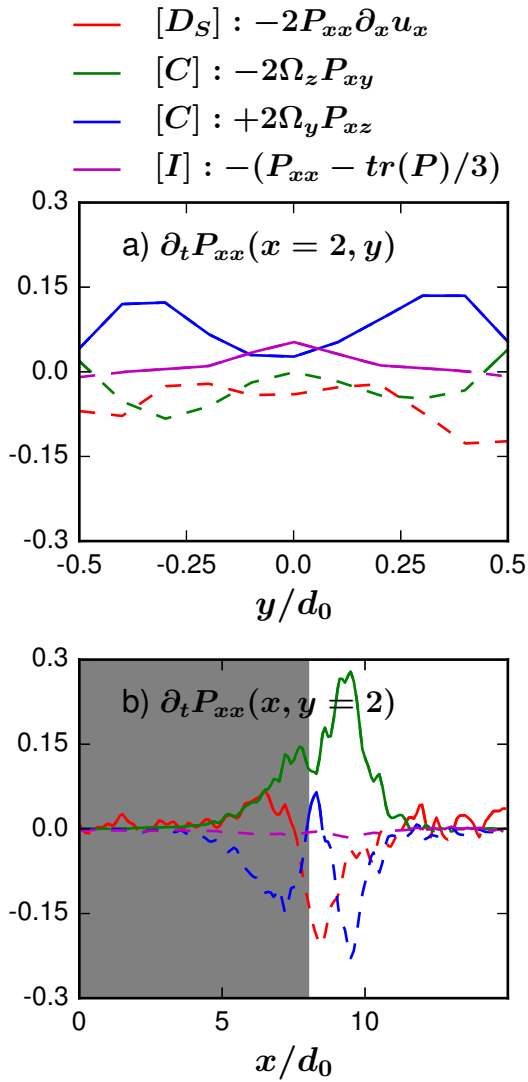}
\caption{Run B. Cuts for the terms defined in Eq.~(\ref{eq_final}) for $P_{xx}$ at $t$ = 80 $\Omega_0^{-1}$: (a) crossing electron layer in $Y$ direction (at $X$ = 2 $d_0$); (b) crossing separatrices in $X$ direction (at $Y$ = 2 $d_0$), grey area represents above the separatrices region. The format for coloured lines is the same as in Fig.~\ref{fig:pyy}}\label{fig:pxx}
\end{figure}

Fig.~\ref{fig:pxx} (a) shows the contribution of the terms from Eq.~(\ref{eq_final}) for $P_{xx}$ in the electron layer. The isotropization term $[I]$ is clearly dominant in the mid-plane $Y$ = 0. While for $P_{yy}$ the term $[I]$ decreases $P_{yy}$, it increases $P_{xx}$ and also $P_{zz}$, although not shown here. For all components considered, this highlights the isotropization nature of this term.

Now focusing at the separatrices, Fig.~\ref{fig:diag_offd_runB} (c) exhibits a clear structure of the enhanced pressure on both sides of the separatrices for the $P_{xx}$ component. It appears from panel (b) of Fig.~\ref{fig:pxx} that this structure results from the term $[C]$ counterbalanced by $[D_{\mathrm S}]$. In the driver part, the gradient of the outflow velocity dominates around separatrices because of the plasma acceleration. The cyclotron part consists of two contributors: $P_{xy}$ increases $P_{xx}$ while $P_{xz}$ decreases $P_{xx}$. For an unbounded cyclotron part growing case without using any isotropization, the counterbalance between $P_{xy}$ and $P_{xz}$ must be limited by the driver, which in turn is not efficient enough on electron time scales.

As a partial conclusion, an electron layer is developing in the mid-plane because of the symmetric part of the driver term for the $P_{yy}$ component. The two other diagonal components are then affected by this growth because of the isotropization term. This growth is not numerically a problem, as it is limited whatever the $\tau$ value in the isotropization term. The picture is quite different at the separatrices: the $P_{xx}$ component is growing essentially because of the $P_{xy}$ which acts through $[C]$. Without any isotropization of both $P_{xx}$ and $P_{xy}$ at the separatrices, these two components would grow in an unbounded way.

We then focus on the sources of the off-diagonal components of the pressure tensor. Fig.~\ref{fig:diag_offd_runB} (d, f) displays the three off-diagonal components at $t$ = 80 $\Omega_0^{-1}$ for run B. The first important point is that the maximum values taken by the off-diagonal components of the pressure tensor are about one order of magnitude smaller than those of the diagonal ones, as already pointed out by other studies of this kind\cite{yin2001, wang2015}. The second point to notice is that the patterns we observe are located in the electron layer and at the separatrices. The electron pressure agyrotropy is observed even where electrons are magnetized. Enhanced non-gyrotropy has been highlighted in separatrix regions in 2D full-PIC simulations~\cite{scudder2008}, and can easily be understood as this region is precisely the topological boundary between the upstream (cold) and downstream (heated) electron populations.

Because of the very low values of the off-diagonal components of the pressure tensor, the analysis of their origin is susceptible to noise. We focus on the $P_{xy}$ component as we will show that it is the important term, both in the electron layer (with physical consequences) and at the separatrices (with numerical issues). As a general structure, $P_{xy}$ has a quadrupolar structure in the electron layer embedded in the second one of opposite polarity at the separatrices.

\begin{figure}
\includegraphics{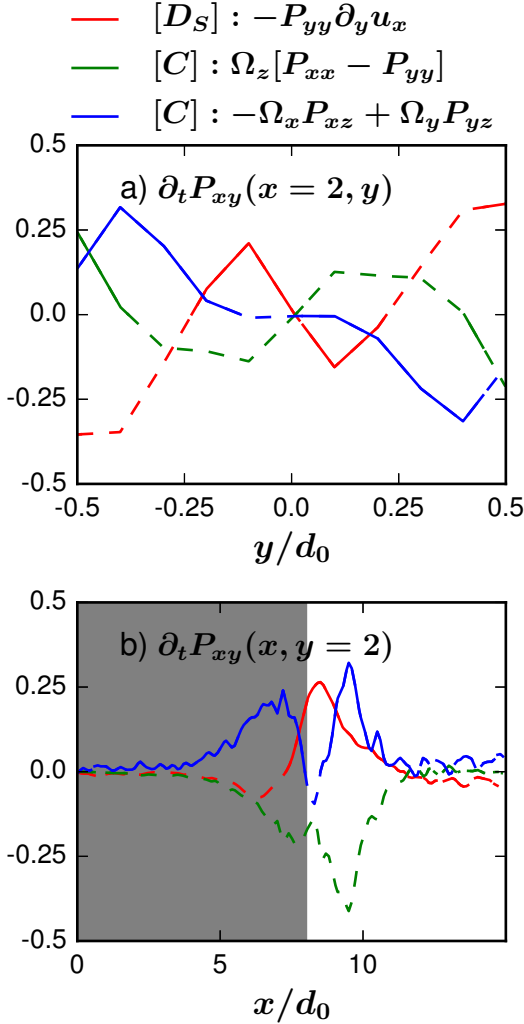}\caption{Run B. Cuts for the terms defined in Eq.~(\ref{eq_final}) for $P_{xy}$ at $t$ = 80 $\Omega_0^{-1}$: (a) crossing electron layer in $Y$ direction (at $X$ = 2 $d_0$); (b) crossing separatrices in $X$ direction (at $Y$ = 2 $d_0$), grey area represents above the separatrices region. The format for coloured lines is the same as in Fig.~\ref{fig:pyy}}\label{fig:pxy}
\end{figure}

In the electron layer, as depicted on panel (a) of Fig.~\ref{fig:pxy}, $P_{xy}$ is growing because of term $[D_{\mathrm{S}}]$ involving the $P_{yy}$ component, and balanced by the term $[C]$ involving anisotropy $P_{xx}-P_{yy}$. The growth of $P_{xy}$ component in the electron layer is then a consequence of the growth of $P_{yy}$ component and is thus observed at the same scales. In the same way as for $P_{yy}$, this growth is bounded by the thickness of the electron layer because of the regularizing effect of the agyrotropic part of the term $[C]$.

We now focus on the structure of $P_{xy}$ component at the separatrices. While not depicted, the effect of  $[D_{\mathrm{C}}]$, $[D_{\mathrm{A}}]$ and $[I]$ terms are very marginal, and the $[D_{\mathrm{S}}]$ and $[C]$ terms seem to cancel out each other. But the structure at the separatrices is twofold: the value of $P_{xy}$ is vanishing at the separatrices but is observable at its inner and outer edges. On panel (b) of Fig.~\ref{fig:pxy}, using the same format as Fig.~\ref{fig:pxx} (b), we observe that the term $[C]$ is there at play, as well as the term $[D_{\mathrm{S}}]$ involving the $P_{yy}$ component and the gradient in the $Y$ direction of the velocity in the $X$ direction. Without any isotropization, the $P_{xy}$ component at the separatrices is growing up to unrealistic values because of the cyclotron term involving mainly $P_{xz}$. Hence, the counterbalance for the $P_{xx}$ component involving both $P_{xy}$ and $P_{xz}$ is weak without the isotropization effect of the divergence of the heat flux.

As a partial conclusion, $P_{xy}$ is growing in the electron layer because of the driver, and is limited by the anisotropy $P_{xx}-P_{yy}$. There, no matter how weak the isotropization term is, these structures are not problematic from a numerical point of view. At the separatrices, the growth of $P_{xy}$ is more sensitive as it results from the growth of $P_{yy}$ (associated with the anisotropy), and the off-diagonal component $P_{xz}$, which in turn is growing because of the anisotropy $P_{xx}-P_{zz}$.

A general picture can now be drawn in the electron layer :
\begin{itemize}
\item a thin structure of enhanced value of $P_{yy}$ is developing at electron scale in the very middle of the current sheet.
\item this structure also develops for $P_{xx}$ and $P_{zz}$ because of the isotropization term.
\item  this resulting electron layer also exhibits a $P_{xy}$ component developing as a direct consequence of the anisotropy $P_{xx}-P_{yy}$.
\end{itemize}
as well as at the separatrices :
\begin{itemize}
\item  the $P_{xx}$ and $P_{xy}$ components are mutually feeding one the other, essentially through the term $[C]$.
\item  if not limited by the isotropization term, these two components of the pressure tensor grow in a numerically unstable way.
\end{itemize}

\section{Macroscopic consequences for the current sheet} \label{sec:runCD}

As concluded in the previous section, the growth of $P_{xx}$ (and the associated anisotropy $P_{xx} - P_{zz}$) as well as the growth of $P_{xy}$ at the separatrices are key points in the numerical stability of the system. The isotropization term, modeling the role of the divergence of the electron heat flux, acts in a way to limit the growth of both $P_{xx}$ and $P_{xy}$. Several numerical attempts clearly showed that only for these two terms, the isotropization was mandatory for numerical stability. Runs C and D are intended to evaluate the consequences of the isotropization term efficiency on $P_{xx}$ and $P_{xy}$.

The reason for run C is to evaluate how the six-component electron pressure tensor evolves if the isotropization term only acts on $P_{xx}$ and $P_{xy}$ components. The reason for run D is less obvious and needs some words: $P_{xy}$ happens to play a key role in the time evolution of both $P_{xx}$ and $P_{yy}$. But the summation of the equations governing their time evolution cancels out the term involving $P_{xy}$, meaning that run D is the one where the destabilizing effect of $P_{xy}$ is artificially removed.

\begin{figure}
\includegraphics{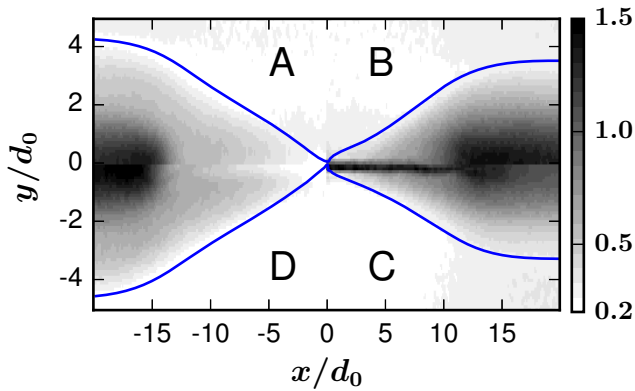}
\caption{ $P_{yy}$ at $t$ = 80 $\Omega_0^{-1}$ normalized to the initial maximum value. The four quadrants are depicting the results coming from the four runs given in Tab.\ref{tab}: $A$ \textit{(top left)}, $B$ \textit{(top right)}, $C$ \textit{(lower right)} and $D$ \textit{(lower left)}. The separatrices are depicted in blue line.} \label{fig:pyy_2d}
\end{figure}

To be able to compare the results for runs A, B, C, and D, we display the results for each of these runs in a given quadrant of the ($x,y$) space (taking advantage of the symmetry in both $X$ and $Y$ directions). Fig.~\ref{fig:pyy_2d} displays for these four runs the $P_{yy}$ component normalized to the initial maximum value. The electron layer is clearly visible in run B and even more pronounced in run C with a longer extent in the $X$ direction. But this pressure structure is missing in run D, meaning that the appearance of this electron layer critically depends on the growth of the $P_{xy}$ component.

As detailed in the previous section, $P_{yy}$ is built-up because of the $[D_{\mathrm{S}}]$ term which is free of the electron mass. But the growth of this term is limited by the cyclotron term which depends on the electron mass as well as the $B_x$ component of the magnetic field. This highlights the fact that the thickness of the field reversal is also an important scale playing a role in defining the width of this electron layer. These arguments are then totally coherent with the width of the electron layer, which is at sub-ion scale, but not totally controlled by the electron scales.

\begin{figure}
\includegraphics{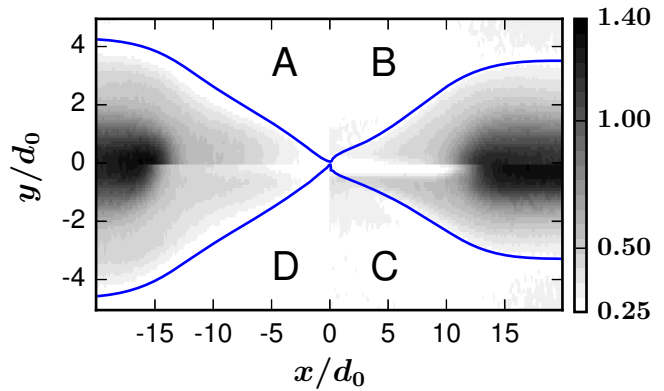}
\caption{Electron density at $t$ = 80 $\Omega_0^{-1}$ normalized to the initial maximum value, for runs A \textit{(top left)}, B \textit{(top right)}, C \textit{(lower right)} and D \textit{(lower left)}. The separatrices are depicted in blue line.} \label{fig:dens}
\end{figure}

\begin{figure}
\includegraphics{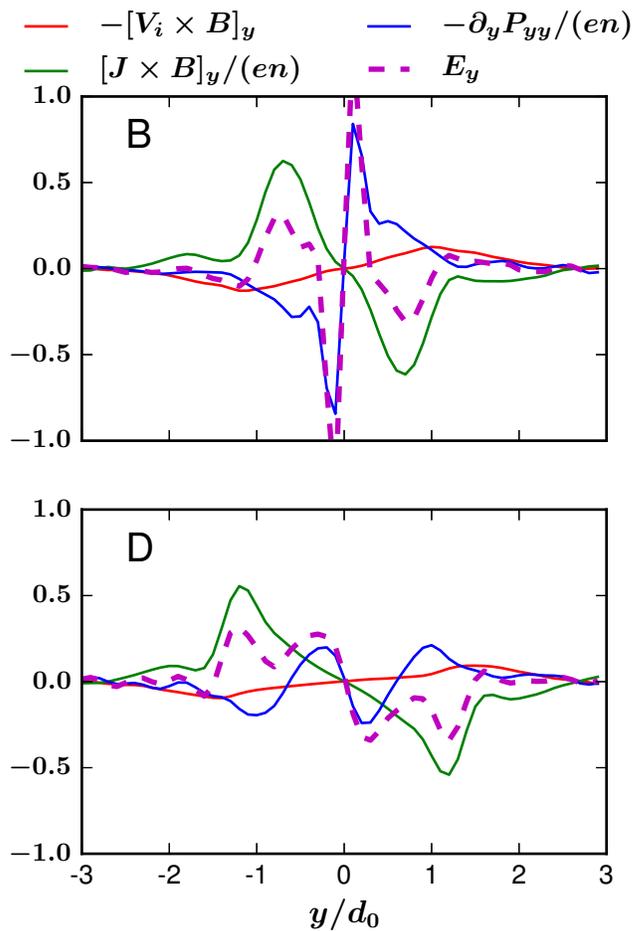}
\caption{Cut in $Y$ direction at $X$ = 4 of the Ohm's law terms Eq.~(\ref{ohm}) for $E_y$ component (averaged over 10 time steps for 79.5$<$ $t$ $<$ 80.5): run B with layer (upper panel), run D without layer (lower panel).} \label{fig:ey_ohm}
\end{figure}

The $P_{yy}$ structure can interestingly be compared to the density given in Fig.~\ref{fig:dens}. For run B and even more clearly for run C, one observes a density gap in the electron layer, between the two density bulges downstream from the X-point. We then need to investigate the $Y$ component of the electric field (coming from the electron Ohm's law) in order to understand why the ions are expelled out from the electron layer. All the terms of the Ohm's law are depicted in Fig.~\ref{fig:ey_ohm}, except the one involving the gradient of $P_{xy}$ and the hyper-viscous dissipative term as their values are negligible.

For run B, the $E_y$ electric field has a very sharp bipolar structure embedded in this electron layer which gets the particles out of the electron layer. This is the main difference from run D, where such a bipolar gradient pressure term exists, but with a smaller amplitude and an opposite direction. Such layer of the bipolar electric field embedded in a thicker layer of the oppositely bipolar Hall electric field has been already studied by full-PIC simulations~\cite{chen2011} and has been attributed to the meandering orbits of the electrons. In the presented hybrid modeling, the shape of the electric field is mainly defined by the gradient of the $P_{yy}$ component in the $Y$ direction, such an increase of $P_{yy}$ is compatible with heating usually resulting from the bounce motion of the electrons. The estimated bounce scale~\cite{hesse2011} in the mid-plane vicinity is 0.2$d_0$ for run B and 0.5$d_0$ for run C at t = 80$\Omega_{0}^{-1}$, that is comparable with the electron layer width. The evolution of the pressure tensor in a fluid model does not truly capture the meandering electrons, but at least it allows to capture the consequences of the electron heating. The density factor amplifies the pressure structure as density has a local minimum in the mid-plane, meaning that both the pressure gradient and the density profile are responsible for the strongly peaked associated electric field. Inspecting the terms for the time evolution of $P_{yy}$ displayed in Fig.~\ref{fig:pyy}, none of them is associated with any density gradients, meaning that there is no mechanism that could limit the growth of $P_{yy}$ and the associated density gap. Furthermore, these structures stay quite stationary, allowing the force balance analysis (Section~\ref{sec:layer}).

\begin{figure}
\includegraphics{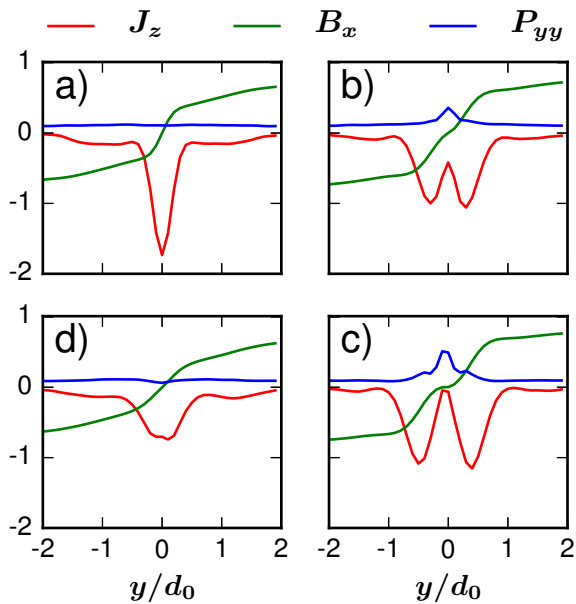}
\caption{ Cut in $Y$ direction at $X$ = 0 (averaged over 10 time steps for 79.5$<$ $t$ $<$ 80.5) : out-of-plane current $J_z$ (red), magnetic field $B_x$ (green), diagonal pressure component $P_{yy}$ (blue). For a) run A, b) run B, c) run C and d) run D.} \label{fig:jz_Cut}
\end{figure}

Fig.~\ref{fig:jz_Cut} shows a cut through the X-point for the out-of-plane current $J_z$. We observe for run A the standard pattern: the scalar pressure is somewhat constant through the current sheet and the reversal of magnetic field is quite sharp in the middle of the current sheet. As a consequence, the associated $J_z$ current has a highly peaked structure.

For runs B and C, the situation is quite different. Because of the increased $P_{yy}$ component in the electron layer, the magnetic field inside the electron layer has a lower value compared to run A. To maintain this plateau in the magnetic field, the associated current $J_z$ is also smaller at the very middle of the electron layer. This straightforwardly results in a bifurcated structure of the current sheet that we observe at the sub-ion scale. The minimum of current in the mid-plane is even more pronounced for run C, which outlines the clear role of the development of the $P_{xx}-P_{yy}$ anisotropy and $P_{xy}$ agyrotropy. This conclusion is consolidated by run D, where the current sheet is no more bifurcated, remembering that this run is the one where these two terms are artificially quenched. The current density bifurcation that we observe is corresponding to the reduction of the out-of-plane electron flow velocity in the mid-plane, which has already been observed in previous studies~\cite{matsui2008}.

\begin{figure}
\includegraphics{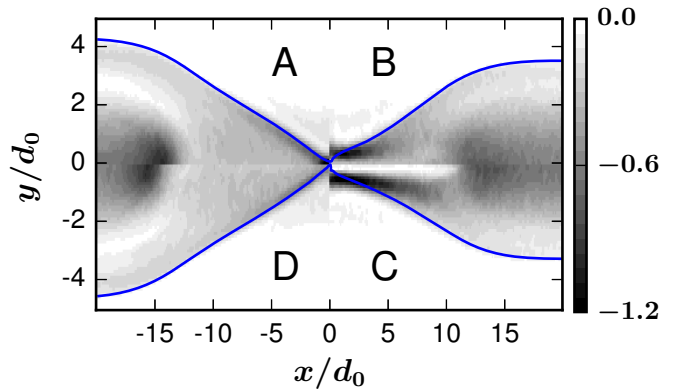}
\caption{Color-coded value of the $J_z$ component of the total current at $t$ = 80 $\Omega_0^{-1}$, for runs A \textit{(top left)}, B \textit{(top right)}, C \textit{(lower right)} and D \textit{(lower left)}. The separatrices are depicted in blue line.} \label{fig:jz}
\end{figure}

Fig.~\ref{fig:jz} displays the two-dimensional spatial configuration of the $J_z$ component of the total current. As commonly observed in 2D reconnection studies, for run A and D, the thickness of the out-of-plane current sheet is diminished close to the X-point, and advected downstream. The picture is quite different for runs B and C: this current is totally vanishing at $Y$ = 0, but expelled above and below the mid-plane, resulting in a bifurcated current sheet. Such a bifurcated current sheet (BCS) has already been discussed theoretically~\cite{zelenyi2003}, numerically~\cite{delcourt2004} and with in-situ data observations~\cite{runov2003}. Observed numerically~\cite{daughton2004, matsui2008, liu2014} BCSs were carried by non-gyrotropic electron pressure, associated with anisotropic heating, due to electrons heated in the direction perpendicular to the magnetic field.

On the numerical side, using 2D full-PIC simulations, one obtains an extended current layer~\cite{shay2007}, which thickness decreases approaching the realistic ion to electron mass ratio. Out from the reconnection context, BCS has also been reported, associated with the Lower-Hybrid-Drift-Instability developing in the current sheet\cite{daughton2004} and supposed to be a natural way for the current sheet evolution~\cite{karimabadi2005}. It is suggested~\cite{liu2014} that the BCS scale depends on the electron bounce width, which depends on the thickness of the current sheet and the electron mass. They hence clearly observe a decrease of BCS size when going to the realistic electron mass, but it is still observed at this limit. As the tearing mode is suggested to be more unstable~\cite{matsui2008}, this gives no insight about the efficiency of the reconnection process. On the observation side, while many in-situ observations of BCS have been reported at ion scale~\cite{hoshino1996, asano2005, runov2006}, a recent study also reported the observations of BCS at electron scale~\cite{norgren2018}.

As pointed out in previous studies~\cite{chen2011}, numerical simulations, resolving the electron inertial length of the magnetic reconnection without guide-field, have shown a double peak structure of the out-of-plane current density. Such structures have been attributed to the electron meandering motion at the magnetic field reversal. But such bifurcated current sheets are not always clearly observable~\cite{ng2011}, which may be related to a failing spatial resolution as already pointed out~\cite{lapenta2007}. Furthermore, a clear picture of their origin is still lacking. For the observed numerically BCS~\cite{hesse2001, shay2007}, the electron mass greatly influences the current sheet structure. As known, increasing ion to electron mass ratio up to realistic values requires a significant reduction of resolution. We also performed B-type runs with decreasing mass ratio $\mu$=10 while keeping the same spatial and temporal resolution. Then, the distance between the two peaks of the out-of-plane current slightly increases while the absolute minimum depth in the mid-plane becomes less pronounced. This results from the fact that the pressure anisotropy is controlled by the regularization effect of the cyclotron term, and that the larger $\mu$, the smaller the scale at which this term is operative. Furthermore, the current magnitude in the mid-plane can be null for the unrealistic C case, while for run B it always keeps a finite value, as generally observed~\cite{runov2006}. But investigation of mass ratio dependence is out of the scope of this paper, and needs further analysis.

It is also important to note the $X$-extent of BCS, on the order of 5 ion inertial lengths for run B. It is even larger (about 10 $d_0$) for run C, which supports the fact that the enhanced value of $P_{yy}$ in the electron layer (not isotropized in run C) is on the origin of this structure. The problems of the physical mechanism for electrons cooling in the mid-plane and the heat flux closure are still unclear for hybrid models with pressure tensor evolution and need for future detailed comparisons with full-PIC simulations to better understand and better model the differences that we observe.

\begin{figure}
\includegraphics{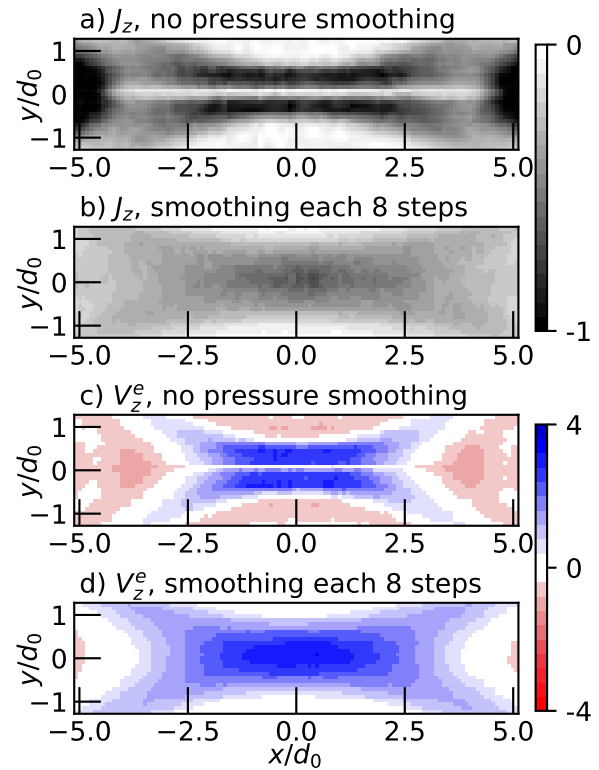}
\caption{Color-coded value of (a, b) the out-of-plane current $J_{z}$ and (c, d) the out-of-plane electron flow velocity $V^e_{z}$ at $t$ = 20 $\Omega_0^{-1}$ for hybrid-PIC runs under the same conditions as the one of Kuznetsova et al. 1998\cite{kuznetsova2001} ($\mu$ = 25, $T_e/T_i$ = 0.2, $\lambda$ = 0.5 $d_0$ and $n_b$ = 0.2). Panels (a, c) show cases without any smoothing on the pressure tensor components and panels (b, d) show the cases with a smoothing (using a three-points stencil) every 8 time steps.} \label{fig:kuz_cur}
\end{figure}

The current structure obtained and discussed in Fig.~\ref{fig:jz} has not been reported in previous similar hybrid-PIC studies\cite{winske1994, hesse1995, yin2001, kuznetsova2001}. In order to understand this discrepancy, Fig.~\ref{fig:kuz_cur} displays the profile of the out-of-plane current under the same conditions as the one of Kuznetsova et al.\cite{kuznetsova2001} with (panel b) and without (panel a) the same smoothing on the pressure as in this previous study. Panels (c, d) of Fig.~\ref{fig:kuz_cur} display the out-of-plane electron flow velocity. It clearly appears that without any smoothing, the electron flow velocity is also bifurcated, following the full-PIC results~\cite{matsui2008}. The bifurcation of the current sheet is wiped out by such a smoothing. While such smoothing can be required for numerical stability, it can have significant consequences in lowering or removing such small-scale structures.

\section{Comparison with full-PIC simulations}
\label{sec:kinetic}
Reconnection lead to bifurcated current sheets in a number of simulations~\cite{camporeale2005}. As follows from the previous considerations, in hybrid-PIC simulations extended by ten-moment description for the electron fluid, one has to avoid even a modest smoothing on the components of the electron pressure tensor to observe the bifurcation development. This has been outlined using the GEM~\cite{kuznetsova2001, birn2001geospace} challenge setup as a case study. While such initialization could look outdated, it appears to be interesting for making comparison, both with numerous previous studies, and with full-PIC simulations.

\begin{figure}
\includegraphics{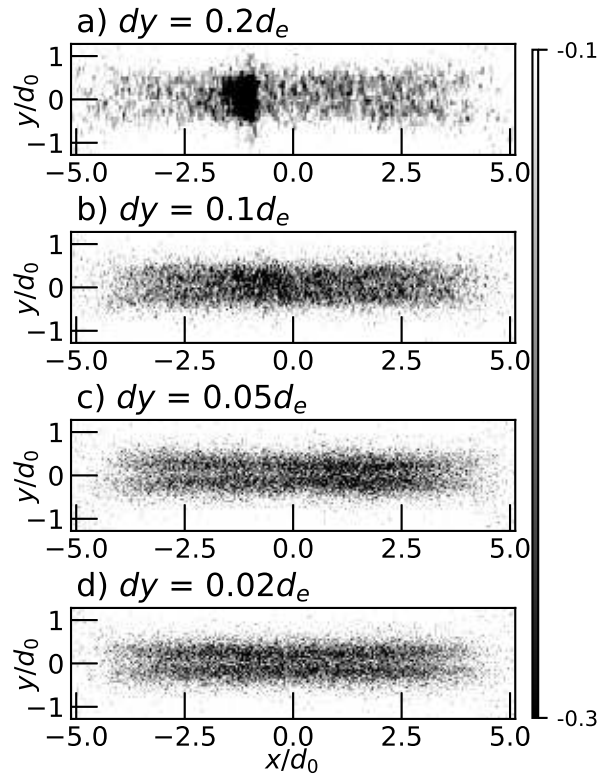}
\caption{ Color-coded value of the out-of-plane current $J_{z}$ at $t$ = 20 $\Omega_0^{-1}$ for full-PIC runs performed using code SMILEI~\cite{derouillat2018}, initial conditions from work of Kuznetsova~\cite{kuznetsova2001} ($\mu$ = 25, $T_e / T_i$ = 0.2, $n_b$ = 0.2$n_0$): (a) grid resolution is 0.2$d_e$, (b) 0.1$d_e$, (c) 0.05$d_e$ and (d) 0.02$d_e$. } \label{fig:smilie}
\end{figure}

As the grid resolution is a keystone for the bifurcation development~\cite{lapenta2007}, we then have carried out a set of four full-PIC runs performed with the code SMILEI~\cite{derouillat2018,smileicode} using a grid cell size from 0.2 down to 0.02 electron inertial length. The other parameters are the same as in the GEM, that is $\mu$ = 25, $T_e/T_i$ = 0.2, $\lambda$ = 0.5 $d_0$ and $n_b$ = 0.2. These four full-PIC runs can then be compared with the hybrid ones~\cite{kuznetsova2001} depicted in Fig.~\ref{fig:kuz_cur}. One can clearly see in Fig.~\ref{fig:smilie} that for a grid size smaller than 0.1 electron inertial length, the current sheet is bifurcated, with a spatial extensions of 5 $d_0$ and 1 $d_0$, in the $X$ and $Y$ directions, respectively. These values are comparable with the ones obtained in the hybrid simulations discussed above.

These results emphasize how such bifurcation is sensitive to the grid size, and the necessity to be small enough to properly resolve the electron scale~\cite{lapenta2007}. In both hybrid-PIC and full-PIC cases, the existence of anisotropy at the field reversal is the cause of the bifurcation. While in the hybrid model the anisotropy is captured by the electron pressure tensor components, a subsequent study would be needed to understand the origin of the anisotropy in full-PIC simulations. More specifically, the meandering motion of the electrons around the center of the current sheet should be addressed to understand the relation between the electron bounce width~\cite{hesse2011} and the scale at which the current sheet is bifurcated.

While the physical origin of a bifurcated current sheet in hybrid-PIC and full-PIC simulations are not necessarily lying on the same mechanism, their development during the fast reconnection process could be an indicative feature of peculiar conditions. A deeper investigation will allow to understand in which conditions this can develop and why. We should also mention that a comparable bifurcation of the out-of-plane current density~\cite{ng2011}, as well as the out-of-plane electron velocity~\cite{wang2015, chen2011}, has already been observed in other full-PIC investigations of the reconnection process while not always deeply discussed.

\begin{figure}
\includegraphics{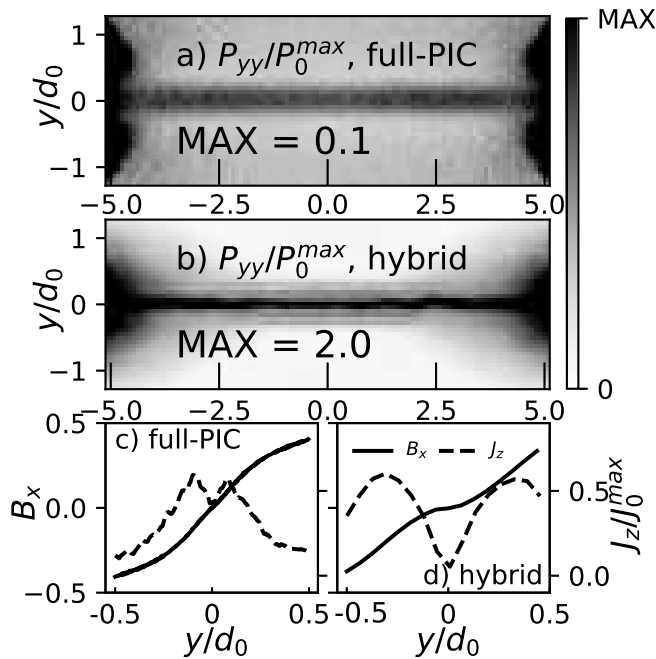}
\caption{ Color-coded value of the $P_{yy}$ component of the electron pressure tensor, normalized to the initial maximum value for (a) full-PIC and (b) hybrid-PIC simulations, at $t$ = 20 $\Omega_0^{-1}$. The lower panels show the anti-parallel magnetic field component (solid lines) and out-of-plane current density (dashed lines) normalized to the initial maximum value for (c) full-PIC and (d) hybrid-PIC simulations.} \label{fig:picvshyb}
\end{figure}

The pressure tensor components in the full-PIC simulations are very similar to the previous one already published~\cite{wang2015}. In panels (a, b) of Fig.~\ref{fig:picvshyb}, we display the most important $P_{yy}$ component for the development of the BCS. The comparison between hybrid-PIC and full-PIC clearly shows that in both cases there is an increase of $P_{yy}$ in the very middle of the electron layer. Panels (c, d) of Fig.~\ref{fig:picvshyb} display the anti-parallel component of the magnetic field and the out-of-plane current density. In both cases the kinetic pressure of the plasma associated with $P_{yy}$ compress the surrounding magnetic field and hence locally decrease the magnitude of the $B_x$ component of the magnetic field (hence forming a plateau) which results in the bifurcation of the current sheet. The scale and amplitude of this bifurcation in the full-PIC case softly differs from the one obtained in the hybrid-PIC case. The comparison between the hybrid-PIC and full-PIC simulations has been done using the GEM-setup, because the associated numerical constraints on these two codes (and formalism) are hardly reconcilable with the parameters of runs A-D (using reasonable CPU time). A grid size of 0.02$d_e$ is an upper limit to observe the BCS with the full-PIC code, while such a very low value is not achievable in hybrid-PIC simulations.

In the full-PIC case, because of the meandering motion of the electrons~\cite{horiuchi1994}, the large dispersion of the electron velocities in the $Y$ direction (mixing the up-going and down-going electrons around the mid-plane) is the cause of the strong increase of $P_{yy}$ at a scale between the electron Larmor radius and the current sheet thickness. This is coherent with the growth of the $P_{yy}$ structure at a sub-ion scale. In the full-PIC case, the electrons also interact with the $E_z$ component of the electric field during their slower ``rotation'' around the small $B_y$ component of the magnetic field~\cite{speiser1965}. Such heating is not considered in the hybrid-PIC cases, and can contribute to explain the differences of the structures displayed in Fig.~\ref{fig:picvshyb}.

\section{Consequences for the reconnection rate} \label{sec:layer}

\begin{figure}
\includegraphics{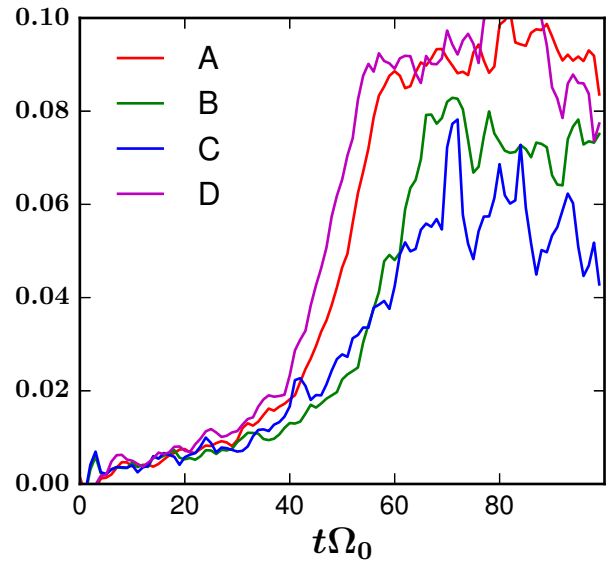}\caption{ $E_z$ component of the electric field at X-point, normalized to the upstream magnetic field and Alfv\'en velocity: red, green, blue and magenta are for runs A, B, C and D, respectively.} \label{fig:rate}
\end{figure}

Fig.~\ref{fig:rate} displays the time evolution of the reconnection rate. It is computed as the local value of the out-of-plane electric field $E_{z}$ at the X-point, normalized to the upstream magnetic field and Alfv\'en velocity, at $X = 0$ and $Y$ = 6 $d_0$. As classically observed in numerical studies of magnetic reconnection\cite{birn2001geospace}, the reconnection rate grows from zero during a transient phase, and then reaches a constant value hence outlining the stationarity of the reconnection process. This value is close to 0.1 for runs A and D, 0.08 for run B, and 0.06 for run C. While not spectacular, these differences are noticeable, and show that the time integration of the six-component electron pressure tensor results in a decrease of the reconnection process efficiency. These features are in agreement with the fact that for runs B and C, the opening angle of the separatrices is smaller than for runs A and D (see Fig. \ref{fig:pyy_2d}, \ref{fig:dens} and \ref{fig:jz}). While in guide-field reconnection electrons stay magnetized all the time, meaning their motion is adiabatic, the reconnection rate reported by Le et al.\cite{le2016} does not depend on the equation of state of the electrons.

It is important to note that for runs A-D, the linear growth rates are almost the same in the early time stage ($t$ $<$ 40 $\Omega^{-1}_0$). In these simulations, $t \sim 40$ is the pivotal time after which the $P_{yy}$ component of the electron pressure starts to significantly grow (as well as the associated off-diagonal components). This is another argument showing that the decrease of the reconnection rate is associated with the bifurcation of the current sheet.

\begin{figure}
\includegraphics{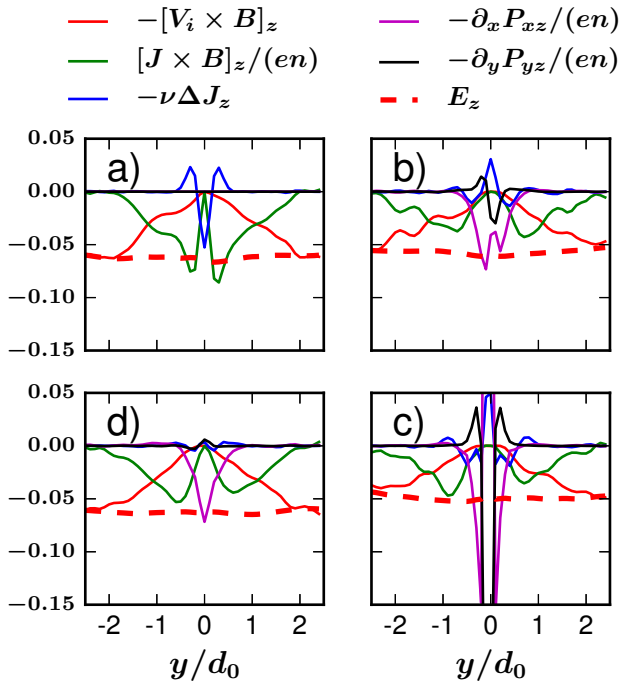}
\caption{Cut in $Y$ direction at $X$ = 0 of the  Ohm's law terms Eq.~(\ref{ohm}) for $E_z$ component (averaged over 10 consecutive output steps for 79.5$<$ $t$ $<$ 80.5). For a) run A, b) run B, c) run C and d) run D.} \label{fig:Ez_cross}
\end{figure}

In Fig.~\ref{fig:Ez_cross}, we display the $E_z$ component of the electric field (thick red dashed line) depending on the $Y$ position for a cut at $X$ = 0. Each term of the Ohm's law is depicted in color, including the two electron pressure terms associated with their agyrotropy.

We observe the classical pattern for run A: in the MHD region, where ions are magnetized ($y > 2 d_0$), the electric field is mainly due to the ideal term associated with the inward advection of plasma in the reconnection region. Closer to the field reversal, the dominant term is the Hall term, associated with the main $B_x$ component of the magnetic field and the $J_y$ component of the current. In the very middle of the current sheet, as $B_x$ vanishes, the hyper-viscous dissipative term is the only one at play, this one being restricted to few grid points. The case of run D is quite similar, except that the Hall term has a smaller amplitude. Furthermore, the pressure term associated with $P_{xz}$ is the leading one in the middle of the current sheet, the hyper-viscous one being negligible. This results from the modified structure of the $J_z$ current and the associated smaller value of its laplacian.

The cases of runs B and C are quite different from A. In Fig.~\ref{fig:Ez_cross}, while dominating in the MHD region, the ideal term is less important as the ion velocity is smaller than in classical case A. A possible mechanism for the reconnection rate reduction is the drag force already discussed by Yin et al.\cite{yin2001}, acting on ions entering the near-X-point region. This drag force could result from the pressure gradient in the inflow direction. In Fig.~\ref{fig:ey_ohm} the bi-polar $E_y$ component of the electric field in the very middle of the electron layer is directed outward from the mid-plane and results from the strong pressure gradient of $P_{yy}$ component. An estimation for the work done by the small-scale electric field embedded in the Hall field is 0.1, which is comparable to the initial kinetic energy of the thermal ions. While the reconnection rate is believed to be determined by ion scale processes, such a highly localized electrostatic field could significantly slow-down the ions convecting through the X-point region. Because of the low plasma compressibility, we also see a net reduction of the ideal term for the reconnection electric field for runs B and C in Fig.~\ref{fig:Ez_cross}. Looking back at Fig.~\ref{fig:dens} we can find an excess of dragged plasma density under and above the X-point region for runs B and C compared to A and D.

Closer to the mid-plane for run B, the reconnection electric field is dominated by the electron pressure term associated with $P_{xz}$. The electron pressure term associated with $P_{yz}$ is also at play, while less important. This pattern is also observable for run C, but as $P_{xz}$ and $P_{yz}$ components are not isotropized, the contribution of these terms for $E_z$ is larger. As the hyper-viscous term is an indicator of the current curvature, one should also notice the term for runs B and C, which has a sign opposite to the one observed in run A. As already pointed out, its form results from the bifurcated nature of the out-of-plane current having an absolute minimum in the mid-plane and a maximum on its edges.

As a partial conclusion of this section, we stress that the reconnection rates reach higher values for the runs without an electron layer (A and D) than for the runs where an electron layer is developing (B and C). The possible mechanism for a smaller out-of-plane electric field is the strongly localized electrostatic field in the mid-plane because of the pressure gradient that slows down the ions convection. The pressure anisotropy in the mid-plane develops because of $P_{yy}$ component, decreasing the relaxation time $\tau$ for the isotropization on that component could reduce the electrostatic field and as result the bifurcation of the out-of-plane current.

\section{Conclusion} \label{sec:discuss}

We studied how the time integration of the electron pressure tensor can be considered in a hybrid code where electrons are treated as a fluid. Aside from an existing implicit method to integrate the fast electron cyclotron part of this tensor, we propose an explicit method based on subcycling. We put forward that such a method converges to the same results as the one obtained with the implicit method, but saving about 30\% of the total computational time. We also outlined the requirement of an isotropization term of this tensor (both diagonal and non-diagonal components) in order to restrain the growth of some of these components, located at the separatrices. Such isotropization is physically associated with the divergence of the electron heat flux. While this term is numerically sensitive to handle, several approximations have been proposed to model it correctly. Such approach is mandatory as the time integration of the exact equation for high order moment (larger than two) can be hazardous from a physical point of view\cite{chust2006} and quite challenging from a numerical point of view. We also show that the $P_{xx}$ and $P_{xy}$ components of the electron pressure tensor are the only ones that need to be isotropized (in Harris-type sheet models) to limit the growth of unstable structures at the separatrices.

We put forward the existence of two characteristic structures in the development of magnetic reconnection in a modified Harris sheet initially pinched: an electron layer at electron scale close to the mid-plane where the diagonal pressure tensor component (dominated by the $P_{yy}$ component) is increased, and an increase of the $P_{xx}$ component at separatrices. These structures are also associated with more complex patterns for the off-diagonal components of the electron pressure tensor. The structures at the separatrices are the most sensitive as their isotropization is mandatory for the numerical stability of the computations. More precisely, $P_{xx}$ and $P_{xy}$ are the components which growth needs to be limited. We show that the $P_{yy}$ structure in the electron layer is associated with an outward $E_y$ electric field, which results in a drag force, acting on ions entering the reconnection region. By force balance, the magnetic field also decreases in the electron layer, which is associated with the splitting of the current sheet. As a consequence, the Hall component of the reconnection electric field is decreased. A clear and direct consequence is that the associated reconnection rate is smaller than in the case of an isothermal closure for electrons where such electron layer is not developing.

We analyzed the different contributions of both physical and numerical effects for the bifurcation of the current sheet, and identified its sources. As the bifurcated current sheet is of a sub-ion scale, a grid size must be small enough and is determined by the given mass ratio. For the full-PIC simulations, the grid size has to be significantly smaller than the electron layer width, which is not the case for the hybrid-PIC simulation. But one also needs to avoid any smoothing on the components of the electron pressure tensor. Such requirement imposes stringent restrictions on the numerical codes where such smoothing can be a living condition.

We performed this study using ion to electron mass ratio $\mu = 100$, which is one order of magnitude smaller than the realistic one. Unfortunately, in hybrid simulations decrease the $\mu$ value would necessitate a smaller grid size, which is resource-consuming in CPU time because of the quadratic relation between time step and grid size. Nonetheless, we conjecture that increasing the mass ratio up to the realistic value would decrease the thickness of the electron layer, and hence increase the gradient of the $P_{yy}$ component. The $E_y$ component of the electric field would then increase, slowing ions on entering the reconnection region.

\section{Acknowledgments}
The research was supported by Center of Exellence "Center of Photonics" funded by The Ministry of Science and Higher Education of the Russian Federation, contract № 075-15-2020-906.  This
research was partially supported by the LPP laboratory. The simulations were partially performed on resources provided by the Joint Supercomputer Center of the Russian Academy of Sciences.

\section{Data Availability}
The data that support the findings of this study are available from the corresponding author upon reasonable request.

\bibliographystyle{apsrev}
\bibliography{biblio.bib}

\end{document}